\documentstyle[prl,twocolumn,aps,epsf]{revtex}

\begin{document}

\title{ \bf Quasicondensation in 2D Interacting Bose Gas: 
Quantum Monte Carlo Study}

\author{Yu.~Kagan$^{1,3}$,V.A.~Kashurnikov$^2$, 
A.V.~Krasavin$^2$, N.V.~Prokof'ev$^1$,
 and B.V.~Svistunov$^{1,3}$}

\address{$^1$ Russian Research Center ``Kurchatov Institute," 123182 Moscow,
  Russia }
\address{$^2$ Moscow State Engineering Physics Institute, 115409 Moscow,
 Russia }
\address{$^3$ Institute for Theoretical Physics, UCSB, 
Santa Barbara, CA 93106}

\maketitle

\begin{abstract} 
We present a detailed Monte Carlo study of
correlations in an interacting two-dimensional Bose gas.
The data for one-particle density matrix in
coordinate representation are compared to the results for the 
local many-particle density correlators, which are responsible
for the recombination rate in real experiments. We found that
the appearance and growth of quasicondensate fluctuations 
changes local correlations between the particles well before 
the Kosterlitz-Thouless transition point.
The amplitude of the $1/m!$-effect is very 
sensitive even to a rather moderate interaction, and is
considerably smaller than its limiting value.
\end{abstract}

\bigskip


The discovery of Bose-Einstein condensation (BEC) in ultra-cold dilute
gases has opened a unique possibility for the study of quantum
correlations in these systems. For one thing, it was observed that 
inelastic processes are suppressed in the presence of BEC
\cite{Burt}, in agreement with theoretical predictions \cite{Kagan85}.
In the 3D gas at low temperature $T \ll T_c$ ($T_c$ is the BEC transition
temperature), the $m$-particle correlator 
\begin{equation}
K_m =n^{-m} \langle \left( \Psi^{\dag } (0,0) \right) ^m  
\left( \Psi (0,0) \right) ^m \rangle \;,
\label{Km}
\end{equation}
is reduced by a factor of $1/m!$ as compared to its value at $T>T_c$
\cite{Kagan85}. Here $\Psi $ is the Bose field operator, and $n$
is the gas density. This result is exact for the ideal gas,
and interaction corrections are small in the gas parameter. In particular,
the rate of three-body recombination (which is proportional to $K_3$) 
must drop by $\sim 6$ times \cite{Kagan85}, as was indeed
measured in Ref.~\onlinecite{Burt}.

In systems with reduced dimensionality (e.g., in $d=2$) at any finite $T$ the
condensate density is zero, and the question of $1/m!$-effect 
is not at all obvious. On another hand, below Kosterlitz-Thouless
 (K-T) temperature $T_c$ 2D system
becomes superfluid due to long-range phase coherence. In the superfluid
state one finds that phase and density correlation
lengths, $R_c$ and $r_c$, may have different scales ($R_c \gg r_c$),
which allows to introduce the notion of quasicondensate 
characterized by the value $n_0$, the
amplitude of the one-particle density matrix $\rho (r)$ at intermediate
distances $ r_c \ll r \ll R_c$ \cite{Kagan87}, where
\begin{equation}
\rho ({\bf r}) =  \langle \Psi^{\dag} ({\bf r}, 0 )
\Psi ({\bf 0},0) \rangle \, .  
\label{rho}
\end{equation}
Local properties of the quasicondensate are identical to those of the genuine
condensate, which justifies the $1/m!$-effect. 
However, in contrast to the 3D case, non-zero $n_0$ in two dimensions
is only due to finite interparticle interaction, which, in its
turn, reduces the drop of $K_m$. Another complication
is associated with the fact that fluctuation region near $T_c$ 
(or $n_c$ if temperature is kept fixed) is rather wide. Also, the 
gas parameter is usually not so small in 2D. Thus, one may observe rather 
broad transformation of local correlation functions near
$T_c$ or $n_c$.

The $1/m!$-effect itself may be used for the detection and study of BEC.
So far experimental attempts to create 2D waveguides for
ultra-cold neutral atoms in specially arranged laser and/or
magnetic fields (see, e.g., Refs.~\cite{Ovchinnikov,Hinds} and
Refs. therein) did not succeed in getting necessary conditions
for BEC. Another promising system is spin-polarized hydrogen on
helium film \cite{Safonov95,Mosk}. Small binding
energy ($\sim 1~K$), and strong delocalization perpendicular to
the film allow to consider the in-plane motion of hydrogen as free. 
Recently it was announced \cite{Safonov98} that the system
undergoes BEC phase transition as the surface density is increased; the
conclusion being based on the significant
drop of the three-particle recombination
rate. The advantage of such measurements in comparison with the standard
search for the K-T transition in
torsion-type experiments \cite{Agnolet} is that they are not
influenced by the substrate and are quasistatic.
 
For realistic interatomic potentials and particle densities  
the fluctuation region is too wide to allow adequate analytic
treatment of correlation functions. We thus attempted 
quantum Monte Carlo (MC) simulation of the 2D Bose gas in the grand
canonical ensemble, by varying the degeneracy parameter $n/(mT)$
(where $m$ is the atom mass) through the chemical potential $\mu$.
Note that experiments with spin-polarized hydrogen on helium
surface are done in the same setup, since surface density is
controlled by $\mu$ of the bulk bufer gas. Surprisingly enough,
under the same conditions the theory of Ref.~\onlinecite{Stoof} 
predicts the decrease of $K_3$ by a factor of $400$! The results
presented below unambiguously prove that the theory of
Ref.~\onlinecite{Stoof} is in error.

Our MC simulation is based on the 
recently developed continuous-time Worm algorithm \cite{Worm},
which is extremely effective in calculating Green functions 
(at any temperature), and is free of systematic errors. Thus
apart from local correlation functions Eq.~(\ref{Km}), we also 
evaluate the one-particle density matrix $\rho (r)$ Eq.~(\ref{rho}). 

We start with the definition of the model Hamiltonian
on the square lattice
\begin{equation}
H = -t \sum_{<ij>} a_i^{\dag }a_j + {U\over 2}\sum_i n_i^2 \;\;,
\;\;n_i = a_i^{\dag }a_i \; \;,
\label{Hub}
\end{equation}
where $a_i^{\dag }$ creates a boson on site $i$, and $<ij>$ stands 
for the pairs of nearest neighbor sites. 
The particular form of the short-range interaction is not important
in the dilute limit, and we restrict ourselves to the on-site Hubbard
repulsion $U$. To ensure that underlying lattice plays no role,
we choose parameters so that characteristic 
one-particle energies are much less than the bandwidth $W=8t$,
i.e., we require $T,U \ll 8t$.
Though the variable ${\bf r}_{ij}={\bf r}_i -{\bf r}_j$
is essentially discrete, in the 
quasi-continuous case, the density matrix 
$\rho (r)$ is a smooth function of \\ $r=\mid {\bf r}_{ij}\mid $.

At the band bottom we may employ quadratic expansion $\epsilon_k=k^2/2m$
for the dispersion law with $m=\hbar^2/(2ta^2)=1/2$ 
(in what follows we use units such that $\hbar =1$, $t=1$ and $a=1$).
The strength of interaction is naturally characterized 
by relative depletion of the condensate density at zero temperature
$\xi = (n-n_0)/n$, which can be easily derived in the framework
of the standard Bogoliubov transformation:
\begin{equation}
\xi = U/8\pi \;.
\label{xi}
\end{equation}
On one hand, we will study the case when $\xi$ is small.
On another hand, the parameter $U$ will be strong enough 
to see the effect of interparticle interaction on local correlators 
$K_m$.

To estimate the critical density we employ 
the universal relation for the K-T
temperature \cite{Nelson}, 
$T_c = \pi n_S^- /2m = \pi n_S^-$,
where $n_S^-$ is the superfluid density at 
$T \to T_c-0$.
At a fixed temperature $T$ the critical value of $n_S^-$ is given by
\begin{equation}
n_S^- =T/\pi \; .
\label{n_S_crit}
\end{equation}
For $T=0.2$ and $T=0.1$ we have $n_s^- \approx 0.064$ and 
$n_s^- \approx 0.032$, correspondingly. Though the 
critical density $n_c > n_S^-$, the difference is normally not large.
Hence, for an estimate of $n_c$ one can use $n_S^-$ (\ref{n_S_crit}).
Note that for both above-mentioned temperatures (for which we will
present below our numerical results) the K-T transition
occurs at relatively low densities.

A typical evolution of $\rho (r)$ with increasing degeneracy 
is presented in Fig.~1. The curves are
obtained for the lattice with $L\times L =80^2$
sites (we use periodic boundary conditions).
In a normal state [Fig.~1(a)] there is only one characteristic 
length-scale - thermal de-Broglie wavelength, 
and $\rho (r)$ decays exponentially  with $r$. 
In Figs.~1(c,d) we see a well-developed quasicondensate state, 
characterized by two different length-scales - 
after fast short-range decrease to a certain value 
(quasicondensate density $n_0$), 
$\rho$ continues to decay very slowly. The case shown in Fig.~1(b) 
is an intermediate one. 
Though there is no  pronounced bimodal shape yet, the decay of $\rho$ at 
larger $r$'s is anomalously slow. This type of behavior is observed
in a rather wide range of variation of the degeneracy 
parameter (fluctuation region).

\vspace{-1.cm}
\begin{figure}
\begin{center}
\leavevmode
\epsfxsize=0.55\textwidth
\epsfbox{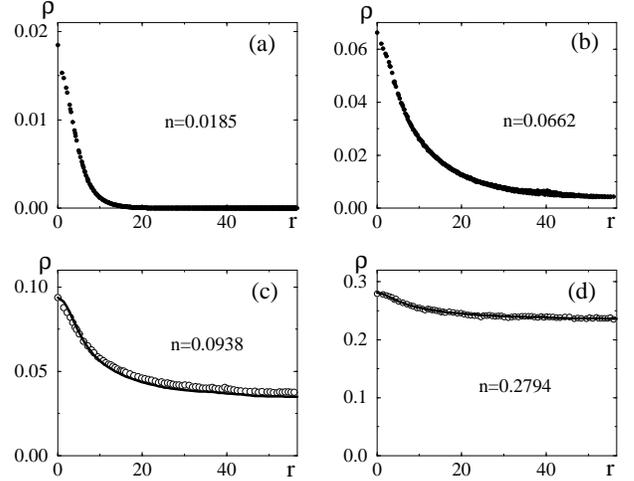}
\end{center}
\caption{Evolution of one-particle density matrix 
with density. $T=0.2$, $U=1.0$. Solid curves in Figs. (c) and (d) are
obtained from  Eqs.~(\ref{rel1}-\ref{U_eff}), and contain { \it no } fitting 
parameters. }
\label{fig:rho}
\end{figure}

Away from fluctuation region (at $n > n_c$) there exists an
analytic representation for $\rho(r)$ \cite{Svi}:
Within Popov's functional-integral approach \cite{Popov} 
one obtains the following set of coupled equations
that has to be solved 
self-consistently for $\rho(r)$, $n_0$, and elementary excitation
spectrum $E(k)= \sqrt{\epsilon (k)[\epsilon (k) + 2n_0 \tilde{U}]}$,
\begin{eqnarray}
\rho (r) & = & e^{-\Lambda (r)} \tilde{\rho}(r) \,, \;\;\;\;
n_0  =  \tilde{\rho}(r \to \infty) \, ; 
\label{rel1}  \\
\Lambda (r) & = & \int \frac{d {\bf k}}{(2\pi )^2} \left[ 1-\cos ({\bf kr}) 
\right]
\frac{\tilde{U}}{E(k)} \nu (k) \,  ,
\label{rel2} \\
\tilde{\rho} (r) & = & n - 
\int \frac{d {\bf k}}{(2\pi )^2} \left[ 1-\cos ({\bf kr}) 
 \right] \nonumber \\
 & & \times \left\{ \frac{\epsilon (k) + n_0 \tilde{U} -E(k)}{2E(k)}
+ \frac{\epsilon (k) \nu (k)}{E(k)} \right\} \,  .
\label{rel3}
\end{eqnarray}
Here $\nu (k)=\{ \exp[E(k)/T]-1 \} ^{-1}$ is the Bose function,
\begin{equation}
\tilde{U} \, = \, \frac{U}{1 + (m U/2\pi ) \ln (1/d_0k_c)} 
\label{U_eff}
\end{equation}
is the effective interaction, $d_0$ is a cutoff for distance, 
and $k_c$ is a typical momentum.
Expression (\ref{U_eff}) implies that $d_0k_c \ll 1$. 
When the degeneracy parameter is on the order of unity or larger, 
to a good approximation one may set in Eq.\ (\ref{U_eff}) $k_c \sim \sqrt{n}$. 
In our model, $d_0$ is just the intersite distance.
The function $\tilde{\rho}(r)$ describes short-range decay 
of the density matrix to the quasicondensate density value $n_0$. The 
long-range decay of $\rho$ is described by slowly growing exponent 
$\Lambda (r)$. For $n > n_c$, we observe a remarkably good agreement 
between $\rho(r)$ calculated from Eqs.~(\ref{rel1}-\ref{U_eff}), 
and our MC results; the agreement becomes 
progressively better away from the fluctuation region
[see Figs.~1(c,d)].

\vspace{-1.cm}
\begin{figure}
\begin{center}
\leavevmode
\epsfxsize=0.5\textwidth
\epsfbox{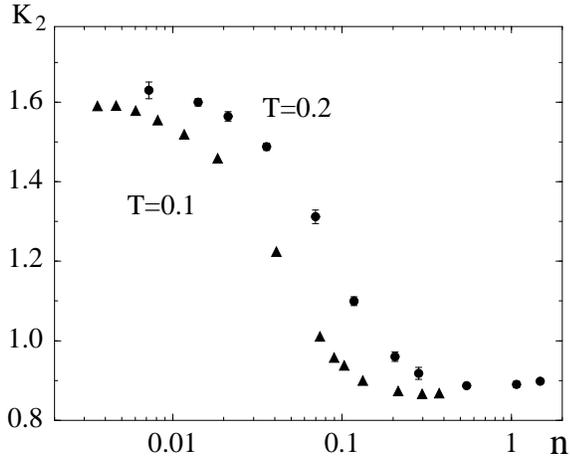}
\end{center}
\caption{The two-particle density correlator as a function of density
at $U=0.4$.}
\label{fig:k2}
\end{figure}
\vspace{-0.6cm}
\begin{figure}
\begin{center}
\leavevmode
\epsfxsize=0.5\textwidth
\epsfbox{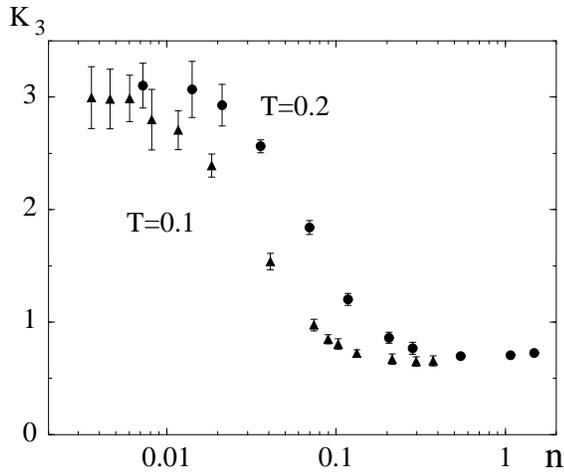}
\end{center}
\caption{The three-particle density correlator as a function of density
at $U=0.4$.}
\label{fig:k3}
\end{figure}
We now turn to the local density correlators (\ref{Km}).
In Figs.\ 2 and 3 we present the data for $K_2$ and $K_3$ as, 
functions of density $n$ for two temperatures
$T=0.2$ and $T=0.1$,  at $U=0.4$.
[The system size ranges from $80^2$, for higher 
$n$'s, up to $300^2$, for lower $n$'s.]
We see a pronounced strong decrease of $K_m$ when
density varies from $n \ll n_c$ to $n \gg n_c$. The most striking result
is the very broad cross-over region. Both correlators start 
to decrease at densities well below $n_c$.

It is important to verify that our results are not artifacts of 
finite-size effects since for finite $L$ 
there exists a considerable fraction of {\it genuine} 
condensate even at {\it finite} temperature. 
(Almost by definition, this fraction is 
given by $\rho(r_*)$, where $r_* \sim L$.) Hence, the
change in local correlators could be due to global condensation 
(like in 3D case), rather than quasicondensation. 
We checked explicitly that the
point $U=1.0$, $n=0.04$, $T=0.2$, 
where there is already a pronounced decrease of
$K_2$ and $K_3$, is 
not sensitive, within the error bars, to the system size for $L=60$,
$100$, and $200$.
Also, we proved that at this point
the value of $\rho(r_*)$ is not a relevant quantity, being
very small at our largest available $L$'s (see Fig.~4).
Most convincingly, in Fig.~4 we see that the 
crossover in $K_2$  starts well before $\rho(r_*)$ becomes appreciable.

\vspace{-0.5cm}
\begin{figure}
\begin{center}
\leavevmode
\epsfxsize=0.5\textwidth
\epsfbox{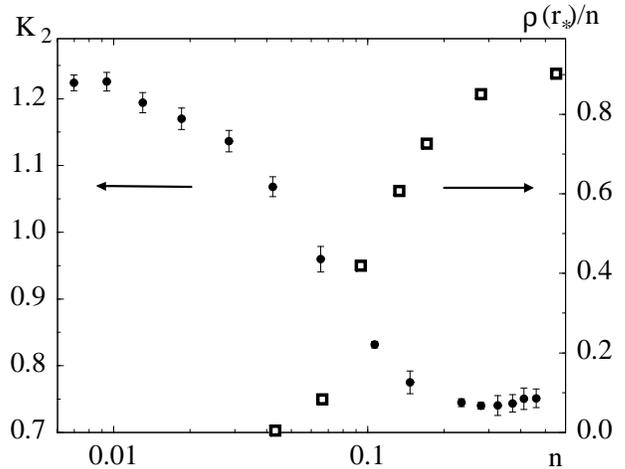}
\end{center}
\caption{$K_2 (n)$ and $\rho(r_*,n)$ curves for 
$U=1.0$, $T=0.2$, demonstrating decoupling of short-range and
long-range correlation properties (here $r_*=40$). }
\label{fig:k2-ro}
\end{figure}

The decrease of $K_m$ in the region  $n<n_c$ 
is indicative of strong quasicondensate fluctuations in the normal state.
In principle, this behavior is not unexpected, since $n_0$ 
is not directly related to the superfluid density $n_S$. 
If the concentration of vortices is small in the fluctuation region
$n <n_c$, the quasicondensate can survive, being related to the 
short-range correlation properties. The variation of $n_0$ thus
has a form of cross-over rather than a transition.

In the limit of very small (but finite) interaction
$K_m$'s should change their values from $m!$ to $1$ 
throughout the transition.
The data presented in Figs.~2 and 3 demonstrate two characteristic
plateaus at $n \ll n_c$ and $n \gg n_c$, the ratio between the two
(at $T=0.1$) being equal to $\approx 4.6$  for
$K_3$, and to $\approx 1.85$ for $K_2$, and
smaller than $m!$. The absolute values of correlators are
also considerably smaller than in the case of negligibly small $U$. 
We thus conclude that even $U=0.4$ is not small enough 
to yeald an idealized picture.
Fig.~5 clearly demonstrates decreasing amplitudes
of the $1/2$-effect with increasing interaction. A similar 
picture was also observed for the correlator $K_3$.
At this point we note that $K(U)$ dependence is 
not universal and can be sensitive to a particular form of
the interaction potential. It is crucial however that 
we observe {\it weakening} of the $1/m!$-effect with increasing $U$, 
which is in a sharp
contradiction with Ref.\ \cite{Stoof} predicting an enormous 
{\it enhancement} of the effect.

\vspace{-1.cm}
\begin{figure}
\begin{center}
\leavevmode
\epsfxsize=0.5\textwidth
\epsfbox{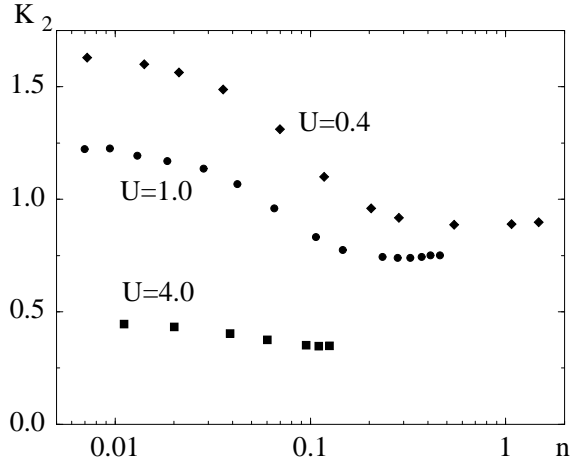}
\end{center}
\caption{$K_2(n)$ curves for various coupling strength $U$, calculated
at $T=0.2$.}
\label{fig:k2-U}
\end{figure}

The three-body dipole recombination rate $W_3$ \cite{Kagan81}
(proportional to $K_3$) of
spin-polarized atomic hydrogen, adsorbed on the surface of superfluid 
helium was measured recently in Ref.~\cite{Safonov98}. 
Hydrogen dynamics perpendicular to the surface is quantized 
since there is only
one localized state, and at relevant temperatures ($\sim 200~mK$)
the system may be considered as purely two-dimensional.
In these experiments $W_3$
was studied as a function of surface
density $n(\mu )$. It was found that recombination rate falls drastically
with increasing $n$, and the decrease of $W_3$ starts well before 
the critical point $n_c$ is reached.
This remarkable result is in qualitative agreement with our MC
simulations. Unfortunately, direct  quantitative
comparison is not possible because hydrogen atom delocalization
perpendicular to the surface is density dependent.
Indeed, $n$ can not exceed some maximum value $n_m$.
When $n \to n_m$ the absorption energy
goes to zero and the hydrogen atom wave-function in the direction 
perpendicular to the surface essentially changes its form 
due to collective effects \cite{KGSS}.
This specific restructuring of the wavefunction leads to the 
additional drop of $W_3$ (through the density-dependent factor
$\alpha(n)$ in $W_3=\alpha K_3$).
As a result, if $n_c$ is close to $n_m$,
the observed rate $W_3$ may drop by a factor much larger 
than $6$ (up to $40$) \cite{KGSS}. 
As far as we can see, this is the only possible
explanation for the measured ratio $W_3(n \ll n_c) /W_3(n \gg n_c) 
> 6$ \cite{Safonov98}.

Summarizing, we studied quasicondensation in a two-dimensional interacting 
Bose system in a rather interesting and experimentally important
regime when interaction is not very small. We traced the evolution
of one-particle density matrix and local correlators with increasing
degeneracy parameter. We found that quasicondensate features appear 
far away from the K-T
transition point. The effect of qusicondensation 
on local correlation properties -- the main 
phenomenon directly relevant to the experiment -- is clearly seen 
in this region, but its strength is rather sensitive 
to the interparticle interaction.

This work was supported by the Russian Foundation for Basic
Research (under Grant No. 98-02-16262) and by the Grant 
INTAS-97-0972 [of the European Community].


\begin{references}

\bibitem{Burt} E.A. Burt, R.W. Christ, C.J. Myatt, M.J. Holland,
E.A. Cornell, and C.E. Wieman,
Phys. Rev. Lett. {\bf 79}, 337 (1997).

\bibitem{Kagan85} Yu. Kagan, B.V. Svistunov, and G.V. Shlyapnikov, 
Sov. Phys. - JETP Lett. {\bf 42}, 209 (1985).

\bibitem{Kagan87} Yu. Kagan, B.V. Svistunov, and G.V. Shlyapnikov, 
Sov. Phys. - JETP {\bf 66}, 314 (1987).

\bibitem{Ovchinnikov} Yu.B. Ovchinnikov, I. Mlnek, and R. Grimm,
Phys. Rev. Lett. {\bf 79}, 2225 (1997).

\bibitem{Hinds} E.A. Hinds, M.G. Boshier, and I.G. Hughes,
Phys. Rev. Lett. {\bf 80}, 645 (1998).

\bibitem{Safonov95} A.I. Safonov, S.A. Vasilyev, I.S. Yasnikov, 
I.I. Lukashevich, and S. Jaakkola,
Sov. Phys. - JETP Lett. {\bf 61}, 1032 (1995).

\bibitem{Mosk} A.P. Mosk, P.W.H. Pinkse, M.W. Reynolds, T.W. Hijmans,
and J.T.M. Walraven, J. Low Temp. Phys. {\bf 110}, 199 (1998).

\bibitem{Safonov98} A.I. Safonov, S.A. Vasilyev, I.V. Yasnikov,
I.I. Lukashevich, and S. Jaakkola, to be published in Phys. Rev. Lett.
 
\bibitem{Agnolet} G. Agnolet, D.F. McQueeney, and J.D. Reppy
Phys. Rev. B, {\bf 39}, 8034 (1989).

\bibitem{Worm} N.V. Prokof'ev, B.V. Svistunov, and I.S. Tupitsyn,
Phys. Lett. A {\bf 238}, 253 (1998); 
Sov. Phys. - JETP, {\bf 87}, 310 (1998).

\bibitem{Stoof} H.T.C. Stoof and M. Bijlsma, Phys. Rev. E,
{\bf 47}, 939 (1993); Physica B, {\bf 194-196}, 909 (1994).

\bibitem{Nelson} D.R. Nelson and J.M. Kosterlitz, 
Phys. Rev. Lett. {\bf 39}, 1201 (1977).

\bibitem{Svi} B.V. Svistunov, Ph. D. Thesis, Kurchatov Institute, 
Moscow, 1990.

\bibitem{Popov} V.N. Popov, {\it Functional Integrals in 
Quantum Field Theory and Statistical Physics}, 
(Reidel, Dordrecht, 1983), Chap. 6.

\bibitem{Kagan81} Yu. Kagan, I.A. Vartan'yants, and G.V. Shlyapnikov,
Sov. Phys. - JETP, {\bf 54}, 590 (1981).

\bibitem{KGSS} Yu. Kagan, N.A. Glukhov, B.V. Svistunov, and 
G.V. Shlyapnikov, Phys. Lett. A {\bf 135}, 219 (1989).



\end{references}
\end{document}